\documentclass[sigconf]{acmart}

\usepackage{amsmath,amsfonts,bm}

\def\eqref#1{equation~\ref{#1}}

\def\1{\bm{1}}

\def\mP{{\bm{P}}}

\DeclareMathAlphabet{\mathsfit}{\encodingdefault}{\sfdefault}{m}{sl}
\SetMathAlphabet{\mathsfit}{bold}{\encodingdefault}{\sfdefault}{bx}{n}

\def\gA{{\mathcal{A}}}

\def\gC{{\mathcal{C}}}
\def\gD{{\mathcal{D}}}

\def\gH{{\mathcal{H}}}
\def\gI{{\mathcal{I}}}

\def\gL{{\mathcal{L}}}

\def\gS{{\mathcal{S}}}

\def\gU{{\mathcal{U}}}

\newcommand{\E}{\mathbb{E}}

\usepackage{subcaption}

\setcopyright{none}
\copyrightyear{}
\acmYear{}
\acmDOI{}

\acmConference[]{}
\acmBooktitle{}
\acmPrice{}
\acmISBN{}

\begin{document}

\title{Recency Dropout for Recurrent Recommender Systems}

\author{Bo Chang, Can Xu, Matthieu Lê, Jingchen Feng, Ya Le, Sriraj Badam, Ed Chi, Minmin Chen}
\affiliation{
    Google, Inc. \\
    Mountain View, CA, USA \\
    \{bochang, canxu, matthieule, jingchenfeng, elainele, srirajdutt, edchi, minminc\}@google.com
}


\begin{abstract}
Recurrent recommender systems have been successful in capturing the temporal dynamics in users' activity trajectories. However, recurrent neural networks (RNNs) are known to have difficulty learning long-term dependencies. As a consequence, RNN-based recommender systems tend to overly focus on short-term user interests. This is referred to as the recency bias, which could negatively affect the long-term user experience as well as the health of the ecosystem. In this paper, we introduce the recency dropout technique, a simple yet effective data augmentation technique to alleviate the recency bias in recurrent recommender systems. We demonstrate the effectiveness of recency dropout in various experimental settings including a simulation study, offline experiments, as well as live experiments on a large-scale industrial recommendation platform.
\end{abstract}




\maketitle

\section{Introduction}

With the ever-expanding corpus of contents across the web applications, recommender systems are increasingly relied upon to help users find the needle in the haystack---to discover the small fraction of items that match users' areas of interest.
The user-item interactions on a platform are naturally sequentially dependent; what the users are interested in next is highly dependent on what they have consumed in the past. In order to surface the right item to the right user at the right time, recommender systems need to have a holistic understanding of the users' areas of interest and preferences, often based on their past activities on the platform. Incorporating sequential information has been shown to improve the performance of recommender systems~\citep{koenigstein2011yahoo,campos2014time,koren2009collaborative,he2016fusing,tang2018personalized,kang2018self,chen2018sequential,yuan2019simple,Garg2019Sequence}.
In particular, recurrent neural networks (RNNs) are widely used to capture the temporal dynamics in users' activity trajectories
\citep{hidasi2015session,beutel2018latent,wu2017recurrent,tang2019towards,chen2019top}.

Despite the successes of recurrent recommender systems, RNNs are known to have difficulty capturing long-term dependencies in the input sequence.
It often fails to utilize input information far into the past due to the vanishing and exploding gradient when trained using backpropagation through time (BPTT)~\citep{pascanu2013difficulty,bengio1994learning}. 
A long line of research has been dedicated to alleviating this issue using techniques such as special initialization schemes and model architecture changes~\citep{hochreiter1997long,cho2014learning,le2015simple,chen18i,arjovsky2016unitary,wisdom2016full,vorontsov2017orthogonality,chang2019antisymmetric}.

The lack of long-term dependencies prompts an RNN-based recommender system to rely heavily on items a user recently interacted with for recommendation and further recommend items that are similar to them. 
In other words, the system overly focuses on users' short-term interests and ignores the long-term ones;
this is referred to as the \emph{recency bias} in this paper.
The bias can cause the recommender systems to pigeonhole users to their most recent interests. 

In this paper, we introduce a data augmentation technique tailored to recurrent recommender systems to better surface users’ long-term interests. 
Data augmentation techniques are widely used in various machine learning fields~\citep{shorten2019survey,fadaee-etal-2017-data,kobayashi-2018-contextual}.
Many of them perform some form of random transformations on the input data.
For example, for image data, the transformations could include random scaling, rotation, flipping, etc.
The extra noise introduced to the input data acts as regularization and makes the model generalize better.

We propose a simple yet effective data augmentation technique named \emph{recency dropout}. It removes the most recent user activities from the input sequence to the recurrent recommender system. 
This in turn forces the model to make use of long-term user interests in the data and mitigates issues caused by the recency bias.

The main contributions of this work include:
\begin{itemize}
    \item We quantitatively study the recency bias in recurrent recommender systems, focusing on a REINFORCE recommender system.
    
    \item We propose recency dropout, a simple yet effective data augmentation technique, to alleviate the recency bias in recurrent recommender systems.
    
    \item Applying the Jacobian analysis, which is used to study the long-term dependencies for RNNs, we illustrate that with recency dropout, the training of RNN becomes better conditioned and the vanishing gradient issue is alleviated.
    
    \item We demonstrate the benefits of the recency dropout technique in large-scale live experiments on a commercial recommendation platform serving billions of users and millions of items.
    
    \item Using the number of daily active users on the platform as a holistic measurement of improved user experience on the platform, we observe that capturing long-term user interests leads to improved long-term user experience.
\end{itemize}

\section{Related Work}

\subsection{Data Augmentation}
\label{sec:related_work_data_aug}

Data augmentation techniques are widely used in various machine learning fields to improve the performance of the model. 
These techniques often artificially enlarge the training dataset by performing random transformations on the input data and act as a form of regularization to alleviate overfitting and improve generalization of the model.

Take image classification tasks in computer vision as an example. 
Before being fed into a neural network, the input images are often transformed by, for instance, horizontally flipping, random cropping, random tilting, and altering the brightness to create multiple instantiations~\citep{shorten2019survey}.
Data augmentations are also commonly used in other fields, such as word substitution in natural language processing~\citep{fadaee-etal-2017-data,kobayashi-2018-contextual}, and randomly masking blocks of frequency channels and blocks of time steps in speech recognition~\citep{Park2019}.

\subsection{Long-Term Dependencies}

Modeling long-term dependencies using RNNs is challenging; the main difficulty arises as the gradient backpropagated through time (BPTT) suffers from exponential growth or decay, a dilemma commonly referred to as the exploding or vanishing gradient~\citep{bengio1994learning,pascanu2013difficulty}.

It has been a long-standing research topic to capture long-term dependencies using RNNs, and various approaches have been proposed.
The gating mechanism is designed to facilitate gradient propagation and is applied in long short-term memory networks (LSTM)~\citep{hochreiter1997long} and gated recurrent units (GRU)~\citep{cho2014learning}.
These models however can still suffer from the same problem of not being able to provably account for long-range dependent patterns in sequences~\citep{belletti2018factorized}.

Identity and orthogonal initialization is another proposed solution to the exploding or vanishing gradient problem~\citep{le2015simple,chen18i}. \citet{arjovsky2016unitary} advocate going beyond initialization and forcing the weight matrices to be orthogonal throughout the entire learning process~\citep{wisdom2016full,vorontsov2017orthogonality}. However, orthogonal weight matrices alone do not prevent exploding and vanishing gradients, due to the nonlinear nature of deep neural networks~\citep{pennington2017resurrecting,gilboa2019dynamical}.

Different from RNNs, the attention and self-attention mechanism can effectively capture long-term dependencies by focusing on the relevant part of the sequence and computing a weighted average of the input~\citep{bahdanau2015neural,vaswani2017attention}. 
They have been successfully applied to sequential recommender systems~\citep{ying2018sequential,kang2018self,tang2019towards}. They, however, do present significant challenges to be utilized in industrial recommender systems with long user histories due to the increased serving cost in the order of $O(M^2)$ compared to $O(M)$ for RNNs, where $M$ is the length of the input sequence.

\subsection{Calibration and Diversity of Recommender Systems}

Recommender systems are known to be subject to a strong feedback loop effect; as the system makes recommendations based heavily on users' recent activities, the user feedback in turn further reinforces the system's recency bias.  
This, as a consequence, can narrow down users' areas of interest and prevent them from exploring other contents available on the platform that are of potential interest. 
Pigeonholing users to a narrow set of interests can negatively affect the users, creating echo chambers or filter bubbles, as well as the ecosystem, missing the opportunities to surface contents from certain providers~\citep{nguyen2014exploring,pardos2020designing,celis2019controlling}.

To mitigate this effect, the notion of calibrated recommendation is proposed, which encourages the distribution of genres or popularity of the recommended items to be calibrated with that of a user's past activities~\citep{steck2018calibrated,kaya2019comparison,zhao2021rabbit}. 
On a high level, it ensures the various areas of interest of a user are reflected proportionally according to his/her historical activities and avoids over-amplifying the main or popular areas of interest of the user.

Another direction is to promote diversity of the recommendation~\citep{Lathia2010Temporal,ziegler2005improving,vargas2011intent}, using methods including determinantal point processes~\citep{chen2018fast,gartrell2016bayesian,wilhelm2018practical} and submodular optimization~\citep{ashkan2014diversified,ashkan2015optimal,qin2013promoting}.
See \citet{kunaver2017diversity} and reference therein for a survey on this research direction.

\section{Background}

In this section, we give an overview of the REINFORCE recommender system~\citep{chen2019top}, and examine the recency bias of the system.

\subsection{REINFORCE Recommender System}
\label{sec:background_REINFORCE}

\begin{figure*}[ht]
  \centering
  \includegraphics[width=0.88\linewidth]{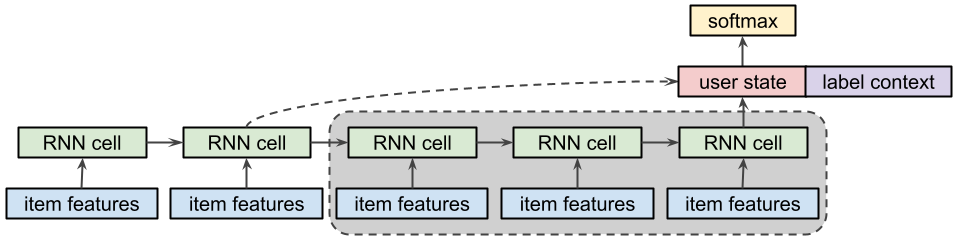}
  \caption{A diagram illustrating the REINFORCE recommender system and the recency dropout technique. The item features of an input sequence $(a_1, \ldots, a_{t-1})$ are passed to an RNN, the last hidden state of which is used as the latent user state representation $s_t$. Together with the label context, a softmax is applied to obtain the policy $\pi_\theta(a_t|s_t)$.
  When recency dropout is applied, the most recent $N$ items are removed from the input sequence, as shown in the grey box in the figure ($N=3$ in this illustrative example). As a result, we use a truncated user activity history $(a_1, \ldots, a_{t-N-1})$ to construct the policy $\pi_\theta(a_t|s_{t-N})$, as denoted by the dashed arrow. This encourages the model to make better use of long-term user interests and alleviates the recency bias.}
  \label{fig:illustration}
\end{figure*}

Let $\gI$ be the item corpus on the platform.
We consider the following sequential recommendation problem: at serving time, given a sequence of a user's historical activities on the platform $\gH_{1:(t-1)} = (a_1, a_2, \ldots, a_{t-1})$, where each item $a_i \in \gI$ is the one that the user interacted with at time $i$, the goal is to recommend a set of items to the user. 
In particular, we return a distribution $\pi\left(a_t \middle| \gH_{1:(t-1)}\right)$ supported on $\gI$; the set of recommended items are generated from this distribution.

We consider the REINFORCE recommender system~\citep{chen2019top}, where the recommendation problem is translated into a Markov decision process (MDP): $\left(
    \gS,
    \gA,
    \mP,
    R,
    \rho_0,
    \gamma
\right)$.
Here $\gS$ is a continuous latent state space describing the user state and context,  $\gA$ is a discrete action space containing items available for recommendation, $\mP: \gS \times \gA \times \gS \rightarrow [0,1]$ is the state transition probability, $R: \gS \times \gA \rightarrow \mathbb{R}$ is the reward function, where $r(s, a)$ is the immediate reward obtained by performing action $a$ at state $s$, $\rho_0$ is the initial state distribution, and $\gamma\ge0$ is the discount factor for future rewards. 

A recommender agent is built by parameterizing and learning a softmax policy
\begin{equation}\label{eq:softmax}
    \pi_\theta(a | s) = \frac{\exp(s^\top v_{a}/T)}{\sum_{a'\in\gA} \exp(s^\top v_{a'}/T)},
\end{equation}
which defines a distribution over the action space $\gA$ conditioning on the user state $s \in \gS$. In our case, the action space is the item corpus $\gA = \gI$. Here, $v_a$ denotes the learned representation for item $a$, and $T>0$ is the temperature that adjusts the entropy of the learned policy. The policy parameters $\theta$ are learned using REINFORCE~\citep{williams1992simple} so as to maximize the expected cumulative reward over the interaction trajectories,
\begin{equation}\label{eq:exp_reward}
\max_{\theta}
\E_{\tau \sim \pi_\theta}
\left[
    R(\tau) 
\right], \mbox{ where } R(\tau) = \sum_{t=0}^{|\tau|} r(s_t, a_t).
\end{equation}
Note that the expectation is taken over the trajectories of $\tau = (s_0, a_0, s_1, \ldots)$ obtained by acting according to the latest learned policy: $s_0\sim\rho_0$, $a_t\sim\pi_\theta(\cdot|s_t)$, and $s_{t+1}\sim \mP(\cdot|s_t, a_t)$.

The recommender agent uses a recurrent neural network (RNN) to model the state transition, i.e., the transition of latent state $s_i$ to $s_{i+1}$ after taking the action of $a_i$.
Therefore, by rolling out the RNN steps, the learned latent state representation at time $t$ can be written as 
\begin{equation}
s_t = f_\theta^{\mathrm{RNN}} \left( \gH_{1:(t-1)} \right) = f_\theta^{\mathrm{RNN}} \left( a_1,\ldots, a_{t-1} \right).
\label{eq:rnn_state_repr}
\end{equation}
Figure~\ref{fig:illustration} shows a diagram of the model architecture of the REINFORCE recommender system.

The algorithm is adapted to the offline (batch) training setup, commonly seen in industrial recommender systems, by applying a technique known as the off-policy correction. 
In particular, the training data $\gD$ consists of users' historical activities in the format of $\left( \gH_{1:(t-1)}, a_t, R_t \right)$ tuples, where $a_t$ is the item consumed by the user at time $t$, and $R_t$ is the discounted cumulative reward associated with $a_t$.
The loss function can be written as
\begin{equation}
\gL = \frac{1}{|\gD|}\sum_{ \left( \gH_{1:(t-1)}, a_t, R_t \right) \in \gD} 
L\left( \gH_{1:(t-1)}, a_t, R_t \right),
\end{equation}
where for each tuple $\left( \gH_{1:(t-1)}, a_t, R_t \right)$, the loss is
\begin{equation}
L \left( \gH_{1:(t-1)}, a_t, R_t \right) = 
\frac{ \pi_\theta (a_t | s_t) }{ \beta (a_t | s_t )}
R_t \log \pi_\theta (a_t | s_t).
\label{eq:loss_function}
\end{equation}
The policy $\pi_\theta$ is defined in Equation~\ref{eq:softmax} and the state representation $s_t$ is computed according to Equation~\ref{eq:rnn_state_repr}.
The behavior policy $ \beta (a|s) $ is an estimate of the historical policy, and the ratio of $ \pi_\theta (a|s) / \beta (a|s)$ is known as the importance weight, which is used to address the data bias caused by the mismatch between the updated policy $\pi_\theta$ and the historical policy $\beta$ that generated the training data.
We refer the interested readers to \citet{chen2019top} for more details about the REINFORCE recommender system.

\subsection{Recency Bias}
\label{sec:background_recency_bias}

The latent state representation is encoded by an RNN as in Equation~\ref{eq:rnn_state_repr}, which is known to have difficulty capturing long-term dependencies in the input sequence. 
As a result, \emph{the recommender system tends to forget about users' activities long in the past and promote contents that are more similar to what a user has recently consumed}. We refer to this as the \emph{recency bias} of the recommender system.

To quantitatively study the recency bias of the REINFORCE recommender system, we examine the degree of ``similarity'' between the policy $\pi_\theta(\cdot|s_t)$ and $a_{t-k}$, the item that the user interacted with $k$ steps ago.
We here measure similarity by the providers and topic clusters\footnote{The topic cluster for each item is produced by: 1) taking the item co-occurrence matrix, where the $(i,j)$-th entry counts the number of times item $i$ and $j$ were interacted by the same user consecutively; 2) performing matrix factorization to generate an embedding for each item; 3) using $k$-means to cluster the learned embeddings into $10,000$ clusters; 4) assigning the nearest cluster to each item. } of the items, assuming items from the same provider or topic cluster are more similar.
More concretely, we investigate the probability mass assigned by the policy $\pi_\theta(\cdot|s_t)$ on items that are uploaded by the same provider of $a_{t-k}$; in other words, we study the following quantity
\begin{multline}
d_{\mathrm{provider}}(k) 
\\ 
= \sum_{a \in \gI} \pi_\theta(a|s_t) \mathbb{I}\{a \text{ and } a_{t-k} \text{ are from the same provider} \}
\label{eq:recency_bias_by_channel}
\end{multline}
as a function of the time difference $k \ge 1$, where $\mathbb{I}$ is the indicator function.
We can similarly define $d_{\mathrm{cluster}}(k)$ based on the topic cluster.
If the functions $d_{\mathrm{provider}}(k)$ and $d_{\mathrm{cluster}}(k)$ quickly decay to zero as $k$ increases, it indicates that the recommender system is strongly recency biased; on the other hand, if the functions are relatively flat, it means that the system is capable of capturing users' short-term as well as long-term interests.

\begin{figure}[htbp]
  \centering
    \begin{subfigure}[t]{0.8\linewidth}
        \includegraphics[width=\textwidth]{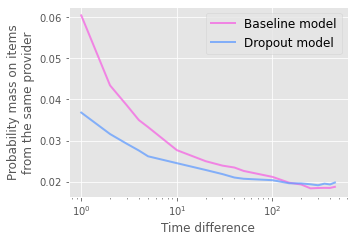}
        \caption{$d_{\mathrm{provider}}(k)$ as a function of $k$.}
        \label{fig:recency_bias_channel}
    \end{subfigure}  
    \\
    \begin{subfigure}[t]{0.8\linewidth}
        \includegraphics[width=\textwidth]{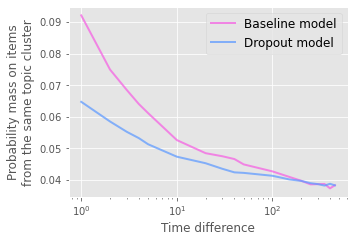}
        \caption{$d_{\mathrm{cluster}}(k)$ as a function of $k$.}     
        \label{fig:recency_bias_cluster}
    \end{subfigure}      
  \caption{Recency bias in the REINFORCE recommender system. It shows the probability mass that the policy $\pi_\theta(\cdot|s_t)$ assigns on items from the same provider or cluster of $a_{t-k}$. On the x-axis is the time difference $k$ on a log scale; on the y-axis are $d_{\mathrm{provider}}(k)$ and $d_{\mathrm{cluster}}(k)$ defined in Equation~\ref{eq:recency_bias_by_channel}.}
  \label{fig:recency_bias}
\end{figure}

In Figures~\ref{fig:recency_bias_channel} and \ref{fig:recency_bias_cluster}, the magenta curves  correspond to $d_{\mathrm{provider}}(k)$ and $d_{\mathrm{cluster}}(k)$, respectively.
It shows that $d_{\mathrm{provider}}(1) \approx 0.06$ and $d_{\mathrm{provider}}(k)<0.02$ for $k>100$; that is, there is a 6\% probability that an item that is similar to what a user most recently interacted with, whereas the probability drops to 2\% for a past item more than 100 time-steps ago. 
The pattern of $d_{\mathrm{cluster}}(k)$ is similar. 
This indicates that there is indeed recency bias in the REINFORCE recommender system.
We will revisit this figure in Section~\ref{sec:offline_expr} to discuss the effect of applying recency dropout.

\section{Recency Dropout}
\label{sec:method}

In this section, we introduce a simple data augmentation or regularization technique named \emph{recency dropout}. It encourages the model to better capture long-term dependencies in users' areas of interest to alleviate the recency bias.

Motivated by other data augmentation techniques discussed in Section~\ref{sec:related_work_data_aug}, the proposed method performs random alterations to the input data.
\emph{During training, we remove the most recent $N$ items from the user's activity history}, where $N$ is a random variable drawn from a distribution supported on non-negative integers $\mathbb{Z}_{\ge 0}$.

Recall that the input data are tuples $\left( \gH_{1:(t-1)}, a_t, R_t \right)$, representing a user's activity history up to time $t$, next action at time $t$, and the associated discounted cumulative reward.
With the proposed recency dropout technique, instead of minimizing $L\left( \gH_{1:(t-1)}, a_t, R_t \right)$ defined in Equation~\ref{eq:loss_function}, we minimize $L\left( \gH_{1:(t-N-1)}, a_t, R_t \right)$.
In other words, part of the user's activity history $(a_{t-N}, \dots, a_{t-1})$ is ablated from the training data. 
This is similar to randomly masking out blocks of the input data proposed by \citet{Park2019}.
Figure~\ref{fig:illustration} illustrates the idea of recency dropout on the REINFORCE recommender system.

Conceptually, we are asking the model to make a recommendation of $a_t$ without using the most recent information, which forces the recommender system to better leverage ``older'' activities $\gH_{1:(t-N-1)}$ instead of relying solely on the most recent ones.

The number of activities to drop $N$ could be sampled from a variety of discrete distributions.
For example, $N$ could follow a discrete uniform distribution on $[N_{\min}, N_{\max}]$. 
Other discrete distributions are also valid choices, including the Poisson distribution, negative binomial distribution, hypergeometric distribution, etc.
We also considered a degenerate distribution as a special case; i.e., $N$ equals a fixed value $N_{\mathrm{fixed}}$ almost surely.

In the same vein with other data augmentation techniques, recency dropout is only applied during training. 
At serving time, the full user activity history $\gH_{1:(t-1)}$ is used as the input to the model.

Note that the proposed recency dropout method should not be confused with the dropout regularization~\citep{srivastava2014dropout} that randomly removes hidden units of a neural network during training, and methods that apply the dropout regularization to RNNs~\citep{gal2016theoretically,semeniuta2016recurrent}. 

\section{Simulation Study}
\label{sec:simulation}

Simulation studies have often been used as a tool to evaluate information retrieval and recommender systems~\citep{yao2021measuring,szlavik2011diversity,carterette2011simulating,diaz2009adaptation}.
To better illustrate the recency bias in recurrent recommender systems, we design and present a simulation study in this section.

The purpose of the simulation is to construct a setting where users have long-term areas of interest that span uniformly among topic clusters, yet a recurrent recommender system tends to focus on recent or short-term interests. Furthermore, we demonstrate how recency dropout effectively mitigates this issue.

\subsection{Simulation Setup}

We notice that in the user activity history we collected on an industrial recommendation platform, consecutive items tend to be more similar: 
there is a 30\% probability that $a_{t-1}$ and $a_t$ are in the same topic cluster and a 26\% probability that them are uploaded by the same provider, where $a_{t-1}$ and $a_t$ are the items a user interacted with consecutively at time $t-1$ and $t$.
Therefore, we design a generative model that simulates sequences of user activities that reflect this property; to that end, we use a Markov chain with a structured transition matrix as the generative model.
We further illustrate that when the learning aims at recovering a user's next interaction based on the activity history, such as the learning objective described in equation~\ref{eq:loss_function}, the recency bias naturally arises.

Let us consider a set of items $\gI$, which can be partitioned into $K$ clusters $\gC_k$ for $k=1,\ldots,K$, such that $\gI = \bigcup_{k=1}^K \gC_k$ and $\gC_k \cap \gC_{k'} = \emptyset$ if $k \neq k'$.
In other words, each item is associated with a unique cluster.
Conceptually, the clusters could represent a notion of users' areas of interest, such as topics, genres, content providers, etc.
We set $\gI = \{0, \ldots, 99\}$ and there are $K=10$ clusters, each has the same number of items. Without loss of generality, we assume $\gC_1 = \{0,\ldots, 9\}, \gC_2 = \{10,\ldots, 19\}, \ldots, \gC_{10} = \{90,\ldots, 99\}$.
We further assume a user's activities follow a discrete-time Markov chain. The transition matrix is denoted by $\mP = (p_{ij}) \in [0,1]^{|\gI| \times |\gI|}$, where $p_{ij}$ describes the probability that the user will interact with item $j$ next given he/she just interacted with item $i$. The transition matrix is designed such that consecutive interactions tend to be on items from the same cluster. 
In particular, the $(i,j)$-th entry of the transition matrix $\mP$ is
\begin{equation}
p_{ij} =
\left\{
  \begin{array}{ll}
    0.7 / 10 = 0.07 & \text{ if $i$ and $j$ are in the same cluster;}  \\
    0.3 / 90 = 0.0033 & \text{ otherwise. }
  \end{array}
\right.
\label{eq:transition_mat}
\end{equation}
It holds that $\sum_{j \in \gI} p_{ij} = 1$ for all $i \in \gI$. 
Note that, for simplicity, we assume a user can repeatedly interact with the same item, which is a valid assumption for some (e.g., music streaming or e-commerce) but not all platforms.

At any time step, the probability that a user stays in the same cluster is $0.7$ and that a user transits to a new cluster is $0.3$. The number of steps it takes for a user to transit to a new cluster follows a geometric distribution with $p=0.3$, the expectation of which is $1/0.3 \approx 3.3$.
Because of the symmetry of the transition matrix, its stationary distribution is a uniform distribution among the item set $\gI$, which can be regarded as the ``long-term'' user interests. 
On the other hand, the most recent item in the activity history can be thought of as the ``short-term'' user interests.
For the next item prediction task, it is apparent that the ``short-term'' user interests are more predictive of the user's next interaction. The proposed recency dropout technique instead encourages the recommender system to recover the ``long-term'' user interests as well. As we will show in the experiment section, capturing long-term user interests leads to improved long-term user experience.

\subsection{Effects of Recency Dropout}
\label{sec:sim_eval_results}

We generate sequences under the simulation setup described above; each has 100 items. A simple sequence model is trained on the generated sequences to predict the last item in the sequence given the 99 interactions before that: the model first performs an embedding lookup for items in the input sequence, then passes the embeddings to a GRU followed by two hidden ReLU layers; finally, softmax is applied to obtain the predictive distribution over $\gI$.
We train the model for 500 steps with a batch size of 128. A separate batch of 1,000 sequences is also generated for evaluation.

\begin{figure}[ht]
    \centering
    \begin{subfigure}[t]{0.9\linewidth}
        \includegraphics[width=\textwidth]{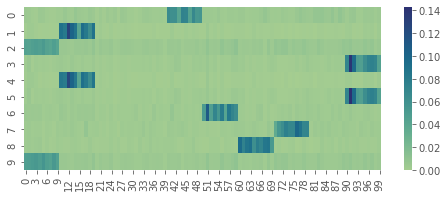}
        \caption{Predictive distribution for the baseline model.}
        \label{fig:sim_pred_heatmap_baseline}
    \end{subfigure}
    \\
    \begin{subfigure}[t]{0.9\linewidth}
        \includegraphics[width=\textwidth]{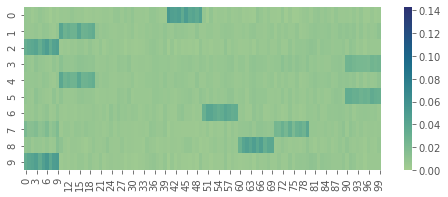}
        \caption{Predictive distribution with recency dropout.}
        \label{fig:sim_pred_heatmap_dropout}
    \end{subfigure}
    \caption{Predictive distributions on simulated data. It shows that the predictive distribution of the baseline model is highly concentrated on items in the same cluster as the last item in the activity history. With recency dropout, the distribution becomes flatter and closer the stationary distribution of the Markov chain, thus better represents users' long-term interests.}
    \label{fig:sim_pred_heatmap}
\end{figure}

We first qualitatively study the effect of recency dropout. 
Figure~\ref{fig:sim_pred_heatmap_baseline} shows an illustrative example of the prediction of the baseline model on 10 evaluation sequences: the predictive distribution is highly concentrated on items in the same cluster as the last item.
As a reference, the last items of the sequences in the batch are $(46, 12,  4, 92, 17, 90, 52, 74, 64, 9)$. 
For example, the first row has the predictive distribution concentrated on items in $[40, 50)$, which are in the same cluster as item 46.
In comparison, we also train a model with recency dropout with $N \sim \gU[0, 5)$.
Figure~\ref{fig:sim_pred_heatmap_dropout} shows the predictive distribution.
It is clear that the distribution becomes flatter and closer to the stationary distribution (uniform) of the Markov chain, thus better represents users' long-term interests.
Meanwhile, it still reflects the short-term interests, but not as strongly as the baseline model.

We also provide a quantitative analysis of the effects of recency dropout.
Two variants are considered: (1) \textbf{random dropout}: $N$ follows a discrete uniform distribution $\gU[0, N_{\max}]$; (2) \textbf{fixed dropout}: $N$ takes a fixed value $N_{\mathrm{fixed}}$.
In order to make the two variants comparable, we report the expected number of dropout for both, that is, $\E(N) = N_{\max} / 2 = N_{\mathrm{fixed}}$.
The evaluation metrics in consideration are as follows: 
\begin{enumerate}
    \item the mean average precision at 1 (mAP@1); 
    \item the mean average precision at 10 (mAP@10); 
    \item the entropy of the predictive distribution;
    \item the Kullback--Leibler (KL) divergence between the stationary distribution (uniform) and the predictive distribution.
\end{enumerate}
Metrics (1) and (2) measure the accuracy of the recommendation; metric (3) indicates the diversity of the predictive distribution; metric (4) measures the ``calibration'' between the predictive distribution and the users' long-term interests~\citep{steck2018calibrated}.

\begin{figure}[ht]
    \centering
    \begin{subfigure}[t]{0.495\linewidth}
        \includegraphics[width=\textwidth]{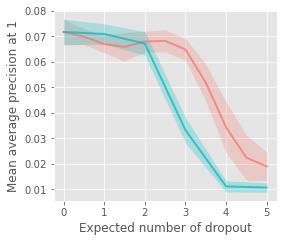}
        \caption{mAP@1.}
        \label{fig:sim_metrics_map1}
    \end{subfigure}
    ~
    \begin{subfigure}[t]{0.495\linewidth}
        \includegraphics[width=\textwidth]{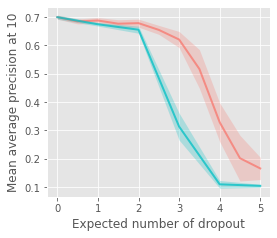}
        \caption{mAP@10.}
        \label{fig:sim_metrics_map10}
    \end{subfigure}
    \\
    \begin{subfigure}[t]{0.495\linewidth}
        \includegraphics[width=\textwidth]{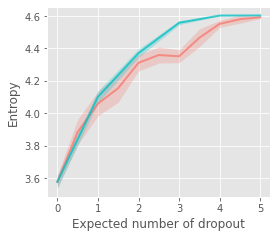}
        \caption{Entropy.}
    \end{subfigure}
    ~
    \begin{subfigure}[t]{0.495\linewidth}
        \includegraphics[width=\textwidth]{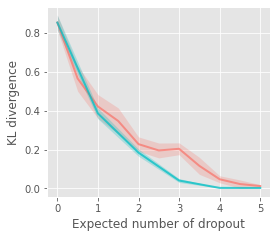}
        \caption{KL divergence.}
        \label{fig:sim_metrics_kl}
    \end{subfigure}
    \caption{Evaluation of two variants of recency dropout on the simulated dataset. The blue curves correspond to the random dropout variant, and red curves correspond to the fixed dropout variant. On the x-axis is the expected number of dropout $\E(N) = N_{\max} / 2 = N_{\mathrm{fixed}}$, and $\E(N)=0$ corresponds to the baseline model without using recency dropout. }
    \label{fig:sim_metrics}
\end{figure}

We evaluate models trained with various $\E(N)$ values; each configuration is trained 10 times with different random seeds. The evaluation batch has 1,000 samples, and the mean and standard error of each metric is reported as a function of $\E(N)$. The model with $\E(N)=0$ is the baseline model without using recency dropout.
Figure~\ref{fig:sim_metrics} shows the results of the quantitative analysis: blue curves correspond to the random dropout variant, and red curves correspond to the fixed dropout variant.
The two variants have similar trends in all metrics. 
Compared with the baseline model, the recency dropout models have slightly worse mAP when the expected number of dropout is small; when it is large, the decrease in mAP becomes more prominent.
Meanwhile, diversity and calibration are improved as the number of dropout increases. 
This indicates that the model has better coverage of users' long-term interests and recommends a more diverse set of items.
In summary, recency dropout improves diversity and calibration, without hurting much of mAP when the number of dropout is relatively small. One can see as the number of fixed dropout $N_\mathrm{fixed}$ reaches $4$ or $5$, the KL divergence shown in Figure~\ref{fig:sim_metrics_kl} reaches 0, suggesting the model recovers the stationary distribution and fully captures users' long-term interests. As explained in the simulation setup, the expected steps it takes for a user to interact with a new cluster is $3.3$, which explains the observed results on the KL divergence.

\begin{figure}[ht]
  \centering
  \includegraphics[width=0.7\linewidth]{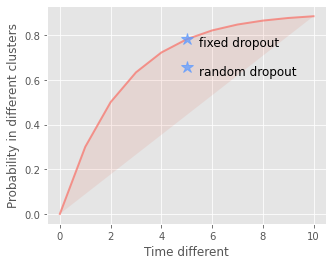}
  \caption{Comparison of random dropout and fixed dropout. The red curve shows the probability that two items that are $k$ steps apart are in different clusters, which can be thought of as a proxy of the difficulty of the task. When comparing fixed dropout with $N_\mathrm{fixed}=5$ against random dropout with $N\sim \gU[0, 10]$. the former has higher expected difficulty by Jensen's inequality, as indicated by the blue stars. For fixed dropout, the expected difficulty is the point on the curve at $k=N_\mathrm{fixed}=5$, whereas for random dropout, it is the center of mass of the curve because $N$ is uniformly distributed.
  }
  \label{fig:random_vs_fixed_dropout}
\end{figure}

Another observation is that, the fixed dropout variant (blue curves in Figures~\ref{fig:sim_metrics_map1} and \ref{fig:sim_metrics_map10}) has lower mAP than random dropout (red curves). 
We argue that this is because the task of fixed dropout is harder than random dropout.
The red curve in Figure~\ref{fig:random_vs_fixed_dropout} denotes the probability that two items that are $k$ steps apart are in different clusters\footnotemark, as a function of $k$.
This can be thought of as the ``difficulty'' of the task when the most recent $k$ items are removed from the input sequence. 
It is evident that this function is concave. 
Using Jensen's inequality, on average, fixed dropout is more difficult than random dropout.

\footnotetext{This probability can be computed recursively. Let $p_k$ be the probability that two items that are $k$ steps apart are in the SAME cluster, and $q_k := 1-p_k$. 
According to Equation~\ref{eq:transition_mat}, we have $p_1=0.7$ and $q_1=0.3$. 
Furthermore, the following recurrence relation holds for $k>1$: $p_k = p_{k-1}p_1 + q_{k-1} q_1 / (K-1)$, where $K=10$ is the total number of clusters. Figure~\ref{fig:random_vs_fixed_dropout} shows $q_k$ as a function of $k$.}

\subsection{Jacobian Analysis}
Improving the trainability of RNN has been a long-standing research topic. Fundamentally, the difficulty is due to the vanishing and exploding gradient issue when training the RNN using backpropagation through time (BPTT) \citep{pascanu2013difficulty}.
Let $h_t$ be the hidden state of an RNN at time $t$, and $1\le t\le T$ where $T$ is the total number of steps.
Training an RNN requires the computation of the gradient of the loss $\gL$ with respect to the hidden state $h_t$:
\begin{equation}
    \frac{\partial \gL}{\partial h_t} = \frac{\partial \gL}{\partial h_T} \prod_{t \le i < T} \frac{\partial h_{i+1}}{\partial h_i}.
\end{equation}
Iteratively multiplying $\partial h_{i+1} / \partial h_i$ causes the gradient to be exponentially vanishing if the largest eigenvalue is less than 1; otherwise, the gradient will exponentially explode. 
This is known as the vanishing and exploding gradient issue, which causes RNNs to have difficulty capturing long-term dependencies in the input sequence. 
Ideally, we want the eigenvalues all to be close to 1 to avoid the vanishing or exploding gradient issue~\citep{chang2019antisymmetric}.

In this section, we study the connection between recency dropout and the eigenvalues of the Jacobian matrix between RNN hidden states.
Intuitively, recency dropout forces the model to leverage information from a long time ago, which is unattainable if the RNN has severe vanishing or exploding gradient issues.

\begin{figure}[ht]
  \centering
  \includegraphics[width=0.7\linewidth]{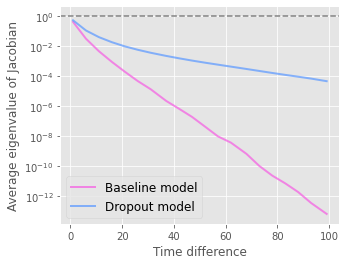}
  \caption{The effect of randomized dropout on mitigating the vanishing gradient issue of RNN. The x-axis represents the time difference $k$ and the y-axis denotes the average eigenvalue of the Jacobian matrix between hidden states that are $k$ steps apart, i.e., $\partial h_{T} / \partial h_{T - k}$, where $T$ is the total number of steps; in this simulation study, $T=100$. The grey dashed line represents the value of 1; the randomized dropout model has eigenvalues that are closer to the grey line, which indicates the vanishing gradient issue is less severe than the baseline model.
  }
  \label{fig:sim_jacobian_eigenvalue}
\end{figure}

Figure~\ref{fig:sim_jacobian_eigenvalue} shows the average eigenvalue of the Jacobian matrix $\partial h_T / \partial h_{T-k}$ computed on a batch of 1,000 evaluation sequences, as a function of $k$. The magenta curve denotes the baseline model and the blue curve corresponds to a recency dropout model with $N\sim \gU[0, 5)$.
Note that the y-axis is on a log scale; visually the two curves are both close to linear, indicating that the gradients vanish exponentially.
Compared with the baseline model, the dropout model has eigenvalues closer to 1, suggesting that recency dropout alleviates the vanishing gradient issue and makes the RNN better capture long-term dependencies in the input sequences.

\section{Offline and Live Experiments}

To measure the effectiveness of recency dropout, we evaluate it offline on a large-scale dataset containing millions of users and tens of millions of items, with billions of feedback between them.
We also verify our approach in live experiments on a commercial recommendation platform serving billions of users.

\subsection{Offline Experiments}
\label{sec:offline_expr}

We extract hundreds of millions of user trajectories from a commercial recommendation platform. 
Each contains a user's historical activity on the platform and the reward, as described in Section~\ref{sec:background_REINFORCE}.
We keep at most 500 historical activities for users with at least one positive interaction.
Among the collected trajectories, 1\% are held out for evaluation. 
The action space or item corpus contains the most popular 10 million items on the platform.
Our goal is to build a recommender system that chooses the next item from the 10 million corpus in order to maximize the cumulative long-term reward for users.

We first demonstrate that recency dropout alleviates the recency bias in the REINFORCE recommender system by revisiting Figure~\ref{fig:recency_bias} in Section~\ref{sec:background_recency_bias}.
Recall that on the y-axis are $d_{\mathrm{provider}}(k)$ and $d_{\mathrm{cluster}}(k)$, measures of how similar the policy $\pi_\theta(\cdot|s_t)$ is from item $a_{t-k}$, which the user interacted with $k$ steps ago, as defined in Equation~\ref{eq:recency_bias_by_channel}; on the x-axis is $k$, the time difference.
The magenta curves correspond to the baseline model without using recency dropout, and the blue curves are the recency dropout model with $N \sim \gU[0, 10)$.
Overall, the dropout model has flatter curves than the baseline model, indicating that the policy captures short-term and long-term interests more evenly.
The difference between the two curves is more prominent when $k < 200$, showing that with recency dropout, the policy no longer focuses too much on recent user interactions. For $k \ge 200$, the dropout model has a (slightly) higher value, which implies that more probability mass is shifted towards longer-term user interests.

For a quantitative evaluation, we use the same set of offline evaluation metrics as in Section~\ref{sec:sim_eval_results}, that is, the mean average precision mAP@1 and mAP@10, the entropy, and the KL divergence. 
Note that the KL divergence is computed between the distribution of the topic clusters in users' past interactions and the topic cluster distribution of the policy. This is to measure how the recommended topics are calibrated with users' existing interests~\citep{steck2018calibrated}.

\begin{figure}[htbp]
    \centering
    \begin{subfigure}[t]{0.495\linewidth}
        \includegraphics[width=\textwidth]{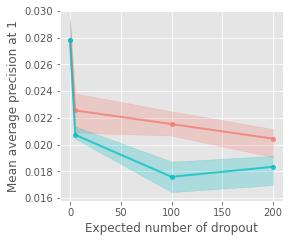}
        \caption{mAP@1.}
    \end{subfigure}
    ~
    \begin{subfigure}[t]{0.495\linewidth}
        \includegraphics[width=\textwidth]{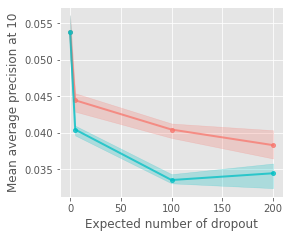}
        \caption{mAP@10.}
    \end{subfigure}
    \\
    \begin{subfigure}[t]{0.495\linewidth}
        \includegraphics[width=\textwidth]{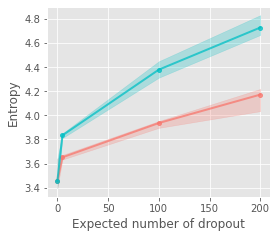}
        \caption{Entropy.}
    \end{subfigure}
    ~
    \begin{subfigure}[t]{0.495\linewidth}
        \includegraphics[width=\textwidth]{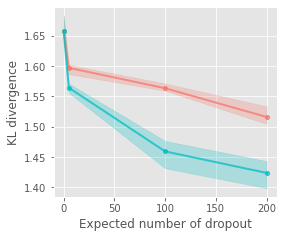}
        \caption{KL divergence.}
    \end{subfigure}
    \caption{Offline evaluation of two variants of recency dropout. The blue curves correspond to the random dropout variant, and the red curves correspond to the fixed dropout variant. On the x-axis is the expected number of dropout $\E(N) = N_{\max} / 2 = N_{\mathrm{fixed}}$, and $\E(N)=0$ corresponds to the baseline model without using recency dropout. }
    \label{fig:offline_metrics}
\end{figure}

Similar to the experiment setup in Section~\ref{sec:sim_eval_results}, we also consider the two variants of recency dropout: random dropout where $N \sim \gU[0, N_{\max}]$ and fixed dropout where $N=N_{\mathrm{fixed}}$. The expected number of dropout is reported $\E(N) = N_{\max} / 2 = N_{\mathrm{fixed}}  \in \{0, 5, 100, 200\}$.
Figure~\ref{fig:offline_metrics} summarizes the evaluation metrics as functions of the expected number of dropout $\E(N)$.
The patterns are similar to that of Figure~\ref{fig:sim_metrics}.
With recency dropout, the mean average precision becomes lower. 
This is expected since the task deviates from the next interaction prediction.
Furthermore, the policy also becomes more diverse, measured by the entropy of the distribution.
Finally, the calibration improves and the policy focuses more on a user's long-term interests.
The difference in metrics between random dropout and fixed dropout is also consistent with that in the simulation study, which suggests that the explanation provided in Figure~\ref{fig:random_vs_fixed_dropout} is plausible.

\subsection{Live Experiments}
\label{sec:live_expr}

\begin{figure*}[htbp]
    \centering
    \begin{subfigure}[t]{0.28\linewidth}
        \includegraphics[width=\textwidth]{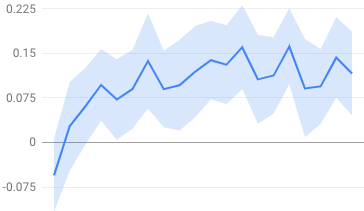}
        \caption{Overall enjoyment.}
    \end{subfigure}
    \quad
    \begin{subfigure}[t]{0.28\linewidth}
        \includegraphics[width=\textwidth]{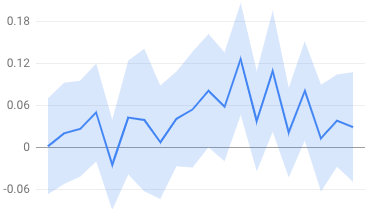}
        \caption{Daily active users.}
    \end{subfigure}
    \quad
    \begin{subfigure}[t]{0.28\linewidth}
        \includegraphics[width=\textwidth]{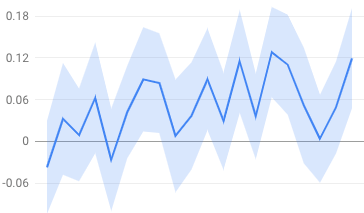}
        \caption{Diversity.}
    \end{subfigure}
    \caption{Live experiment results on the homepage. On the x-axis is the date; on the y-axis is the relative difference in percentage between the experiment and control.}
    \label{fig:live_exp_home}
\end{figure*}

\begin{figure*}[htbp]
    \centering
    \begin{subfigure}[t]{0.28\linewidth}
        \includegraphics[width=\textwidth]{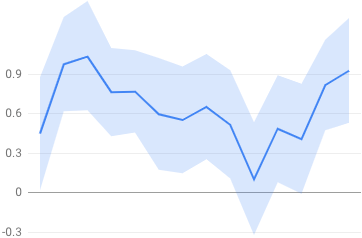}
        \caption{Overall enjoyment.}
    \end{subfigure}
    \quad
    \begin{subfigure}[t]{0.28\linewidth}
        \includegraphics[width=\textwidth]{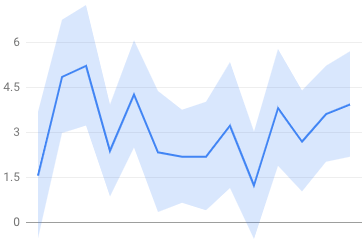}
        \caption{Daily active users with more than 2 hours of activities.}
    \end{subfigure}
    \quad
    \begin{subfigure}[t]{0.28\linewidth}
        \includegraphics[width=\textwidth]{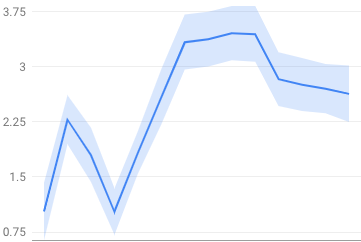}
        \caption{Diversity.}
    \end{subfigure}
    \caption{Live experiment results on the short-form contents service. }
    \label{fig:live_exp_shorts}
\end{figure*}

We conduct a series of A/B experiments in a live system serving billions of users to measure the benefits of the proposed recency dropout technique. The REINFORCE recommender system is built to retrieve hundreds of candidates from a corpus of $10$ million items upon each user request. The retrieved candidates, along with those returned by other sources, are scored and ranked by a separate ranking system before showing the top results to the user. 
Experiments are run for four weeks, during which both the control and experiment models are trained continuously with new interactions and feedback being used as training data. We focus our discussion on metrics capturing: (1) users’ overall enjoyment; (2) the number of daily active users on the platform;
(3) diversity of the user-item interactions, which represents the number of unique topic clusters the users have interacted with.

We first conduct experiments for recommendations on the homepage (browse page) of the app or web page, where users are presented with a whole page of items to choose from.
Figure~\ref{fig:live_exp_home} summarizes the live experiment results using recency dropout with $N\sim \gU[0, 10)$, comparing to a control using the baseline REINFORCE algorithm during the full four-week experiment period. 
On the x-axis is the date, and on the y-axis is the relative difference of a metric in percentage between the experiment and control.
We report the mean and 95\% confidence intervals of the metrics based on the experiment results during the last 7 days.
Relative to the control, the experiment model improves the overall enjoyment by $+0.12\%$ with a 95\% confidence interval of $(+0.08\%, +0.16\%)$.
The number of daily active users also increases by $+0.05\% \ (+0.00\%, +0.09\%)$, which is known to be a difficult metric to increase. 
It indicates an improvement in the long-term user experience.
What is more interesting is that we observe an upward trend in these two metrics during the experiment phase, as shown in Figure~\ref{fig:live_exp_home}, suggesting a user learning effect, i.e., user states change in response to the recommendation policy.
Finally, consistent with the offline experiment results, the diversity metric also improves, leading to a $+0.07\%$ increase with a 95\% confidence interval of $(+0.03\%, +0.11\%)$.

We also conduct a second round of experiments on a new service that provides short-form items to users. Since the contents are much shorter, users are able to interact with many more items within the time frame they spend on the platform. 
To that end, we experiment with much more aggressive dropout; the experiment model applies recency dropout with $N\sim \gU[0, 100)$.
It is again compared against the baseline REINFORCE recommender system as the control.
Figure~\ref{fig:live_exp_shorts} shows the live experiment results on the service of short-form contents. The patterns are similar to that of the experiments on the landing page. The overall enjoyment on this service improves by $+0.56\% \ (+0.26\%, 0.86\%)$; the number of daily active users with more than 2 hours of activities on this service increases by $+2.95\% \ (+2.03\%, +3.89\%)$; and diversity improves by $+3.03\% (+2.75\%, +3.30\%)$. It demonstrates that the benefits of recency dropout are reproducible and transferable across different services.

\section{Conclusion}

In this paper, we investigate the recency bias in recurrent recommender systems, in particular, a REINFORCE recommender system. The recency bias causes the system to recommend items anchoring towards users' short-term interests, ignoring the long-term ones.
In order to alleviate the recency bias and to better reflect users' long-term interests, we introduce the recency dropout technique. 
A simulation study is designed to illustrate the existence of recency bias and to compare the performance of two variants of recency dropout models on simulated data. 
Recency dropout is shown to improve the diversity and calibration of the policy.
The Jacobian analysis also sheds light on how recency dropout improves RNN gradient propagation by alleviating the vanishing gradient problem.
Similar experiment results are obtained through large-scale offline experiments as well.
Finally, we conduct live experiments on an industrial recommender platform serving billions of users and tens of millions of items to verify the benefits and reproducibility of the proposed technique.

\bibliographystyle{ACM-Reference-Format}
\bibliography{arxiv}

\end{document}